\documentclass[%
 reprint,
superscriptaddress,
 amsmath,amssymb,
 aps,
prx,
]{revtex4-1} 

\usepackage{graphicx}
\usepackage{dcolumn}
\usepackage{bm}
\usepackage{afterpage}

\usepackage{color}

\begin{document}


\title{Thermoelectric quantum oscillations and Zeeman splitting in topological Dirac semimetal BaAl$_{4}$}

\author{P. R. Mandal}
\affiliation{Maryland Quantum Materials Center, Department of Physics, University of Maryland, College Park, Maryland 20742, USA}
\author{Kefeng Wang}
\affiliation{Maryland Quantum Materials Center, Department of Physics, University of Maryland, College Park, Maryland 20742, USA}
\author{Tarapada Sarkar}
\affiliation{Maryland Quantum Materials Center, Department of Physics, University of Maryland, College Park, Maryland 20742, USA}
\author{Prathum Saraf}
\affiliation{Maryland Quantum Materials Center, Department of Physics, University of Maryland, College Park, Maryland 20742, USA}
\author{Danila Sokratov}
\affiliation{Maryland Quantum Materials Center, Department of Physics, University of Maryland, College Park, Maryland 20742, USA}

\author{Johnpierre Paglione}
\email{paglione@umd.edu}
\affiliation{Maryland Quantum Materials Center, Department of Physics, University of Maryland, College Park, Maryland 20742, USA}
\affiliation{Canadian Institute for Advanced Research, Toronto, Ontario M5G 1Z8, Canada}

             
\begin{abstract}
Three-dimensional topological semimetals hosting Dirac or Weyl fermions are a new kind of materials class in which conduction and valence bands cross each other. Such materials harbor a nontrivial Berry phase, which is an additional geometrical phase factor arising along the path of an adiabatic surface and can give rise to experimentally measurable quantities such as an anomalous Hall component.
Here we report a systematic study of quantum oscillations of thermoelectric power in single crystals of the topological Dirac nodal-line semimetal BaAl$_{4}$. We show that the thermoelectric power (TEP) is a sensitive probe of the multiple oscillation frequencies in this material, with two of these frequencies shown to originate from the three-dimensional Dirac band. The detected Berry phase provides evidence of the angular dependence and non-trivial state under high magnetic fields. We also have probed the signatures of Zeeman splitting, from which we have extracted the Landé $g$-factor for this system, providing further insight into the non-trivial topology of this family of materials.

\end{abstract}

\pacs{Valid PACS appear here}

\maketitle

\section{Introduction}

The recent discoveries of three-dimensional (3D) topological Dirac and Weyl semimetals \cite{liu2014discovery,yang2015weyl,lv2015observation} having relativistic Dirac or Weyl fermions with exotic properties like high quantum mobility \cite{liang2015ultrahigh,shekhar2015extremely,huang2015observation}, potential topological superconductivity, large magnetoresistance, and symmetry protected linear crossings near the Fermi level \cite{wang2013three, weng2015weyl, huang2015weyl} have stimulated enormous research interest. Among these Dirac materials, the semimetals with non-trivial topological states, topological nodal-line attract much attention due to their particularly interesting physics \cite{fang2015topological}. The examples for these types of materials are ZrSiS \cite{schoop2016dirac}, PtSn$_4$ \cite{wu2016dirac}, PbTaSe$_2$ \cite{bian2016topological}, and HfSiS$_8$ \cite{takane2016dirac} etc., which has been examined by angle-resolved photoemission spectroscopy (ARPES). The unique physical properties of these topological materials provide important information for fundamental theories and they have possible promising applications in optoelectronics, quantum computing, low power spintronics, and green energy harvesting.

Recently, the well-known ``prototype'' compound BaAl$_4$ was identified as a crystalline symmetry-protected topological semimetal with a three-dimensional Dirac spectrum
\cite{wang2020crystalline}. Transport measurements and ARPES data reveal the coexistence of electron and hole pockets in this system, 
with multiple Dirac dispersions creating a saddle-point singularity and the Lifshitz point in between the two Dirac points in momentum space, similar to graphene. In the case of BaAl$_4$, the point group symmetry is C$_{4v}$ on the $\Gamma-Z$ path which is along the rotational axis of the BaAl$_4$ structure. The two doubly-degenerate crossing bands are found to belong to two different representations ($\Gamma6$ and $\Gamma7$) of the symmetry group, and therefore the intersection of the two is protected by the crystalline C$_{4v}$ symmetry, forming a massless Dirac fermion above the Fermi energy \cite{wang2020crystalline}, making it an ideal system to study novel physics of Dirac electrons. Together with Wilson loop calculations, these observations prove the non-trivial topology of the pockets at the Fermi level.

The quantum mechanical Berry phase in materials such as topological insulators, 3D Dirac semimetals, and Weyl semimetals is closely associated to their topological properties \cite{xiao2010berry}, and its non-zero value is a strong indication of the non-trivial topology of electronic structure.
The Berry phase can be determined from the measured phase shift of quantum oscillations \cite{Schoenberg}. The application of a magnetic field causes the cyclotron motion of electrons (that is, closed trajectories in momentum space) and quantization of Landau levels (LLs), which requires a phase change of $2\pi n$, where $n$ is an integer \cite{Schoenberg}. Dirac fermions gives rise a Berry phase of $\pi$, which describes the additional geometrical phase factor for completing a closed orbit in the parameter space. This additional phase factor is a straightforward method for the determination of the Berry phase. 

Here we report observations of temperature- and angular-dependent quantum oscillations in single crystals of the 3D Dirac material BaAl$_4$ using field-dependent measurements of TEP, a sensitive tool for studying fermiology and the topological nature of systems \cite{fauque2013magnetothermoelectric}. The good crystalline quality and sufficiently high electron mobility of BaAl$_4$ \cite{wang2020crystalline}  enables us to observe clear oscillations, allowing for detection of five different frequencies and investigating the phase shifts for two of them. Our results provide further information on the multiple Dirac bands evidenced from the ARPES experiments, and furthermore clearly resolve strong LL splitting at low temperature and high magnetic fields arising from the large Landé $g$-factor of $\sim$ 10.

\begin{figure}[t]
  \centering
\includegraphics[width=80mm]{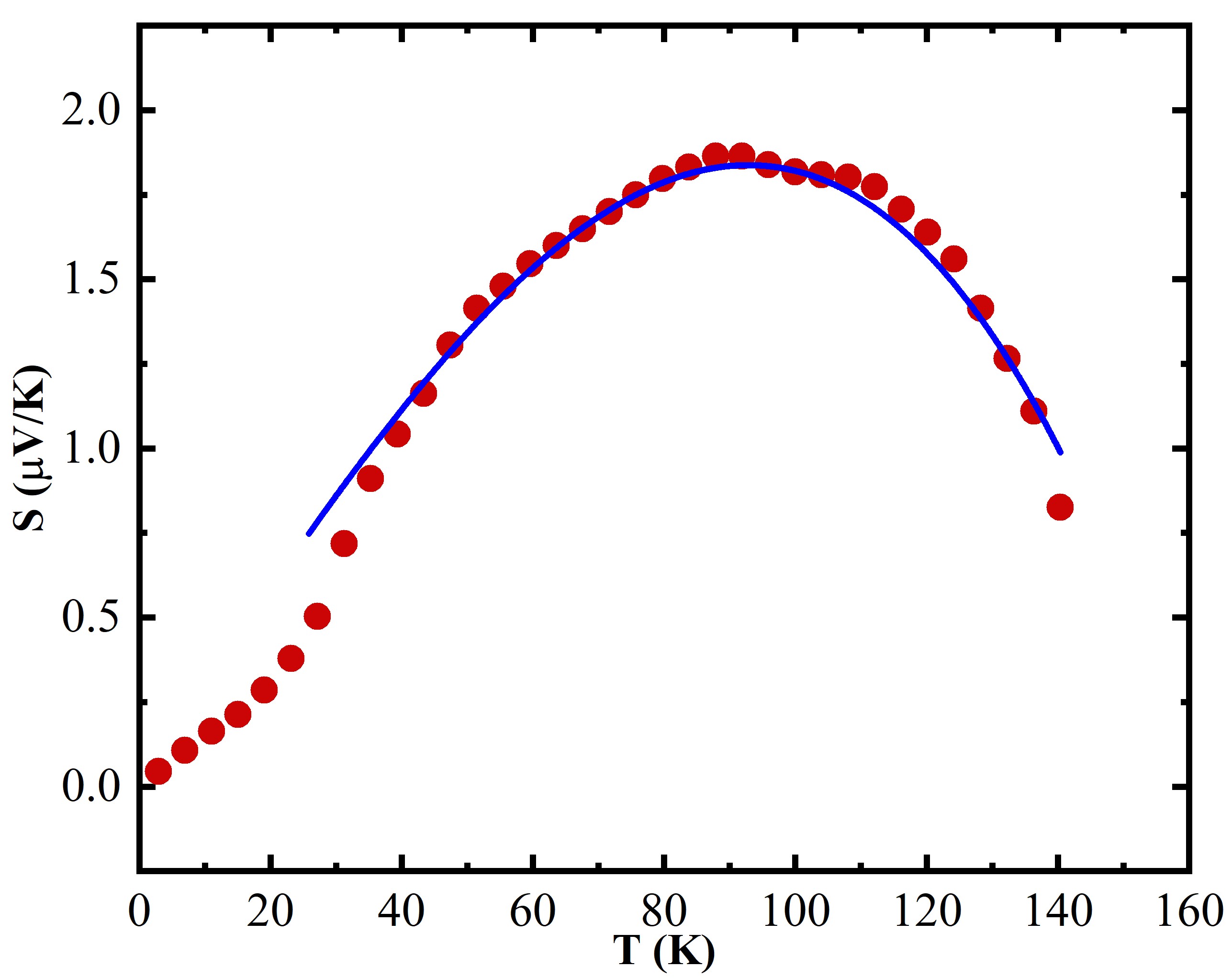}
    \caption{Thermoelectric power of BaAl$_4$ in zero magnetic field. The solid blue line represents the fitting of data according to Eq.~\ref{ST} in the phonon drag region.} 
  \label{Stemp}
\end{figure}

\begin{figure}[t]
  \centering
\includegraphics[width=90mm]{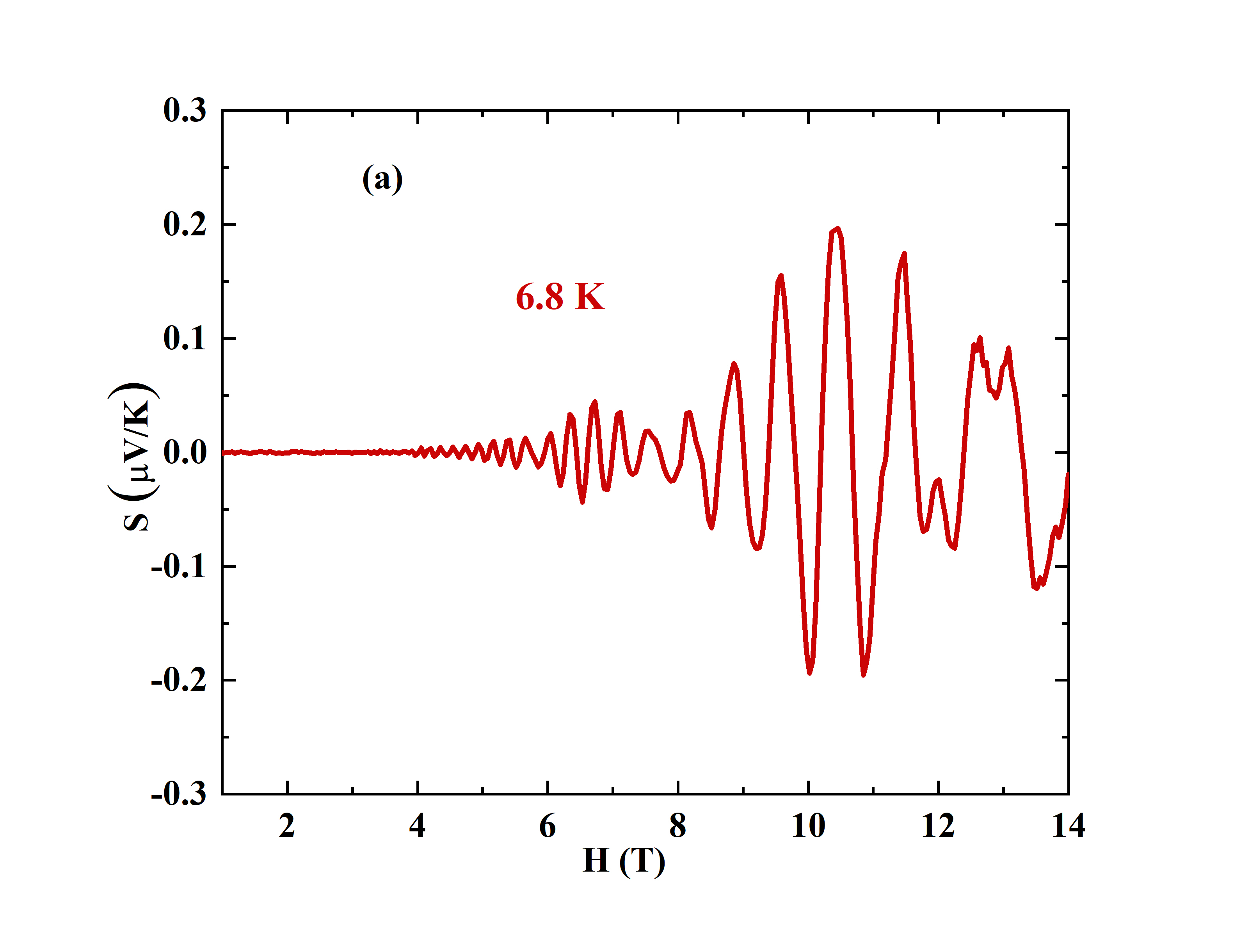}
    \centering
\includegraphics[width=75mm]{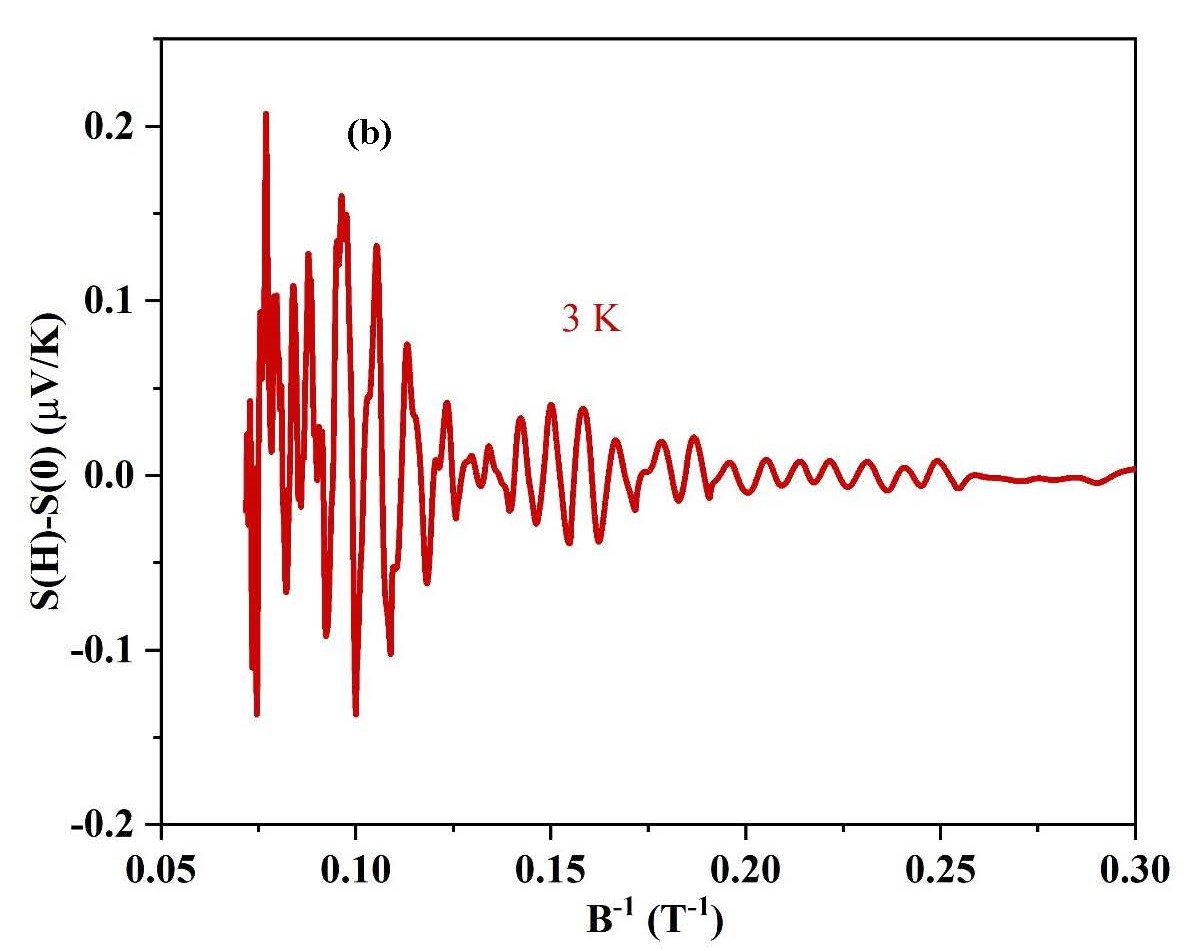}
    \caption{(a) Thermoelectric power as a function of magnetic field for BaAl$_4$ at 6.8 K. (b) The thermoelectric power as a function of inverse magnetic field for BaAl$_4$ at the temperature of 3 K. } 
  \label{Sfield}
\end{figure}

\section{Materials and Methods}

Single crystals of BaAl4 were synthesized by a high-temperature self-flux method. Chunks of Ba (99.98\%, Alfa Aesar) and pieces of Al (99.999\% metal basis, Alfa Aesar) in the ratio of 2:98 were placed in the Canfield alumina crucible set with a decanting frit (LSP ceramics), sealed in the quartz ampule under partial Ar pressure, heated to 1150$^o$C, held at that temperature for 6 hours, cooled down at 3 degree/hour to 950$^o$C with subsequent decanting of the excess Al with the help of a centrifuge. The crystals tend to grow with large faceted surfaces normal to the (001) crystallographic plane. X-ray diffraction data were taken at room temperature with Cu K$\alpha$ ($\lambda$=0.15418 nm) radiation in a powder diffractometer.

Thermopower was measured using the two-heater, two-thermometer method \cite{Thermopower-PRB}, a steady-state technique in which the temperature gradient is periodically reversed to eliminate the Nernst effect and other background contributions. The sample was mounted between two thermally insulated copper blocks attached to a temperature-controlled base. Each block was equipped with a chip resistor heater, while two bare-chip Lakeshore Cernox thermometers were placed at the sample ends to continuously monitor a temperature gradient of 0.7–1 K. The gradient was generated by powering the heaters, and its direction was switched by toggling them on and off. Voltage signals were recorded with a Keithley 2001 multimeter, capable of nanovolt sensitivity, once the gradient stabilized. To minimize random error, multiple measurements were averaged. Phosphor bronze wires, chosen for their negligible thermopower in high fields over the relevant temperature range \cite{Thermopower-highfield}, were used as voltage leads. All measurements were conducted under high vacuum conditions.

\section{Results and Discussion}

As shown in Fig.~\ref{Stemp}, the zero-field thermopower $S$ of BaAl$_4$ shows a non-monotonic temperature dependence and stays positive in the entire temperature range. 
Despite BaAl$_4$ being dubbed a semimetal, $S(T)$ has a form similar to a metal at lower temperatures, increasing with increasing temperature and varying linearly with $T$ up to 20 K, consistent with a bare metallic diffusion thermopower. 
With increasing temperatures above 20 K, S(T) deviates from the linear temperature dependence and develops a peak near 90 K. As shown in Fig.~\ref{Stemp}, data in the range 20--140~K are well fitted to the form 
\begin{equation}
\label{ST}
S = AT+ BT^3,
\end{equation}  
where $A$ and $B$ are fitting parameters representing the contribution from diffusion and the phonon drag scattering process, respectively \cite{GS-Okram}.  This indicates that both diffusion and phonon-drag contributions both contribute in this temperature range. The phonon drag terms are logical here because the phonon drag peak in thermopower is usually observed near temperature $\theta_D/5$, where Debye temperature $\theta_D$ for BaAl$_4$ derivative phases is about 460 K \cite{failamani2016baal} and thus the expected phonon drag contribution maximum matches the observed 90 K peak of our data. The obtained fitting parameters, $A= 3\times10^{-2}~\mu$V/K$^{2}$ and  $B=-1.1\times10^{-6}~\mu$V/K$^{4}$, suggest that charge carrier diffusion thermopower dominates the thermal transport in the temperature range from 20 to 140 K.

\begin{figure}[h!]
  \centering
  \includegraphics[width=80mm]{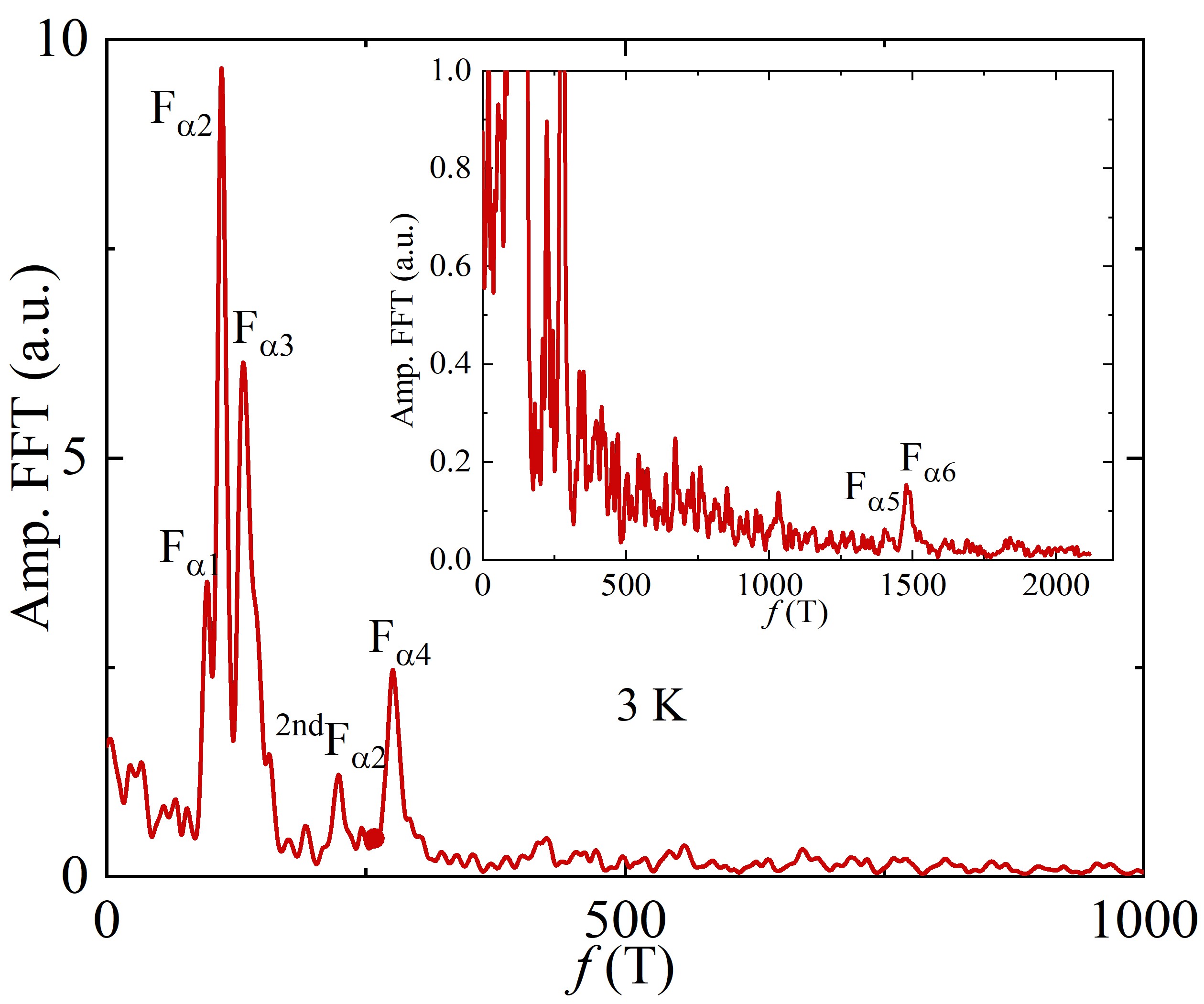}
   \caption{The FFT spectrum calculated for the quantum oscillation of thermoelectric power of BaAl$_4$ at the low frequency region at T=3 K. Inset shows the FFT spectrum in high frequency region at 3K where the peaks are attributed to higher frequencies of $F_{\alpha 4}$ and $F_{\alpha 5}$.  } 
  \label{FFT}
\end{figure}

The thermoelectric response to the oscillations is much more sensitive than the electrical measurements since oscillations in thermoelectric coefficients are driven by changes in entropy and are energy derivatives of the electronic density of states at the Fermi level, as observed for many compounds \cite{trodahl1969quantum, fletcher1981experimental,PRB-Thermopower_QO,QO-ZrSiS}. 
Fig.~\ref{Sfield}(a) represents the thermopower as a function of magnetic field for BaAl$_4$ at 6.8~K. As shown in Fig.~\ref{Sfield}(b), the magnetic field dependence of $S$ for BaAl$_4$ at 3~K readily exhibits quantum oscillations.  
After subtraction of a polynomial background, the oscillations are shown to exhibit several distinct features when plotted as inverse field, which are analyzed using a Fourier transform. Fig.~\ref{FFT} presents the Fourier-transformed frequency spectrum of the 3~K thermopower signal in the large field range from 0 to 1000~T. The observed frequencies for the sample are listed in Table 1, 
and are in good agreement with the previous dHvA experiment for $B\parallel c$ \cite{wang2020crystalline}. 

\begin{table}
\begin{tabular}{ |c|c|c|c|c|c|c| } 
\hline
 & $F_{\alpha1}$ & $F_{\alpha2}$ & $F_{\alpha3}$ & $F_{\alpha4}$ & $F_{\alpha5}$ & $F_{\alpha6}$\\
\hline
frequency (T) & 96 & 111 & 131 & 275 & 1477 & 1490 \\
\hline
{effective mass ($m_0$)} & 0.26 & 0.23 & 0.3 & 0.21 & 0.5 & 0.56\\ 
\hline
\end{tabular}
 \caption{Comparison of effective masses for different bands of BaAl$_4$.}
\label{table}
\end{table}

Because of the exceptional sensitivity of thermoelectric effects to the quantization of the Fermi surface mentioned above, we can observe the $F_{\alpha1}$ oscillation that has not been reported before. In the high-frequency region as shown in the inset of Fig.~\ref{FFT}, the oscillation frequencies $F_{\alpha5} =1477~T$ and $F_{\alpha6} =1490~T$ correspond to the outer branch of the hole pocket centered at the $\Gamma$  point. The small frequency $F_{\alpha4} =275~T$ represents the inner branch of the hole pocket. The two close oscillation frequencies $F_{\alpha3} =131~T$ and $F_{\alpha2} =111~T$ shown in Fig.~\ref{FFT}, correspond to the small hole pockets along the $\Gamma$ and X lines \cite{nakamura2016characteristic}. The high and low-frequency oscillations may originate from the Fermi surface composed of the nodal-line bands \cite{wang2020crystalline}. The extremal Fermi surface cross-sectional areas $A_F$ have been estimated using the Onsager relation $F=(\phi_0 / (2\pi^2 ))A_F$, where $\phi_0= / 2e$  is the magnetic flux quantum, yielding 0.91 nm$^{-2}$, 1.05 nm$^{-2}$, 1.24 nm$^{-2}$, 2.6 nm$^{-2}$, and 14.2 nm$^{-2}$ for  $F_{\alpha1}$, $F_{\alpha2}$, $F_{\alpha3}$, $F_{\alpha4}$, and $F_{\alpha6}$, respectively. 
$F_{\alpha2}$ exhibits a harmonic peak in the FFT spectrum, while the rest of the components do not, which may be due to the large amplitude of its primary frequency as compared to the other oscillations (see Fig.~\ref{FFT}). 

\begin{figure*}[t!]
  \centering
  \includegraphics[width=160mm]{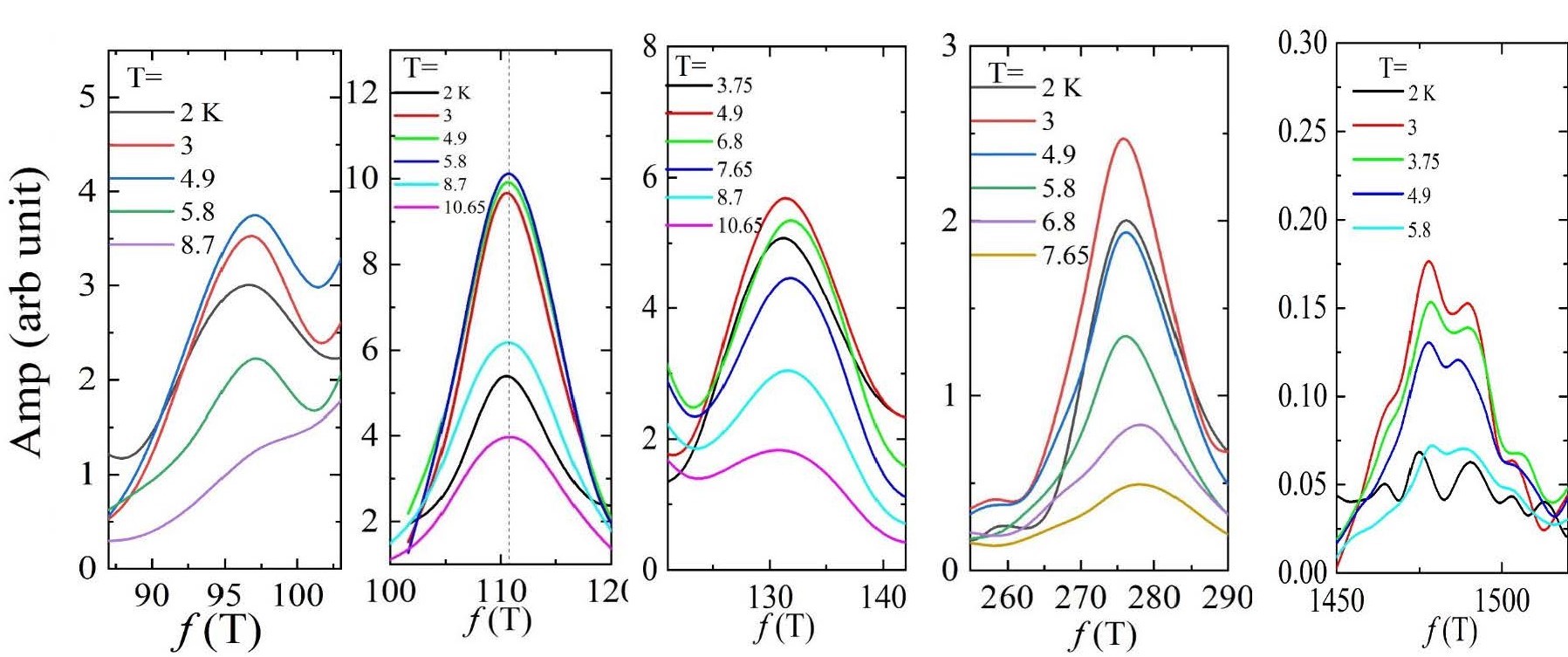}
  \caption {Temperature dependence of FFT peaks obtained from the oscillatory component of the thermoelectric oscillations for $B\parallel c$ calculated at different temperatures.}
  \label{FFTpeaks}
\end{figure*}

\begin{figure*} [t!]
  \centering
  \includegraphics[width=160mm]{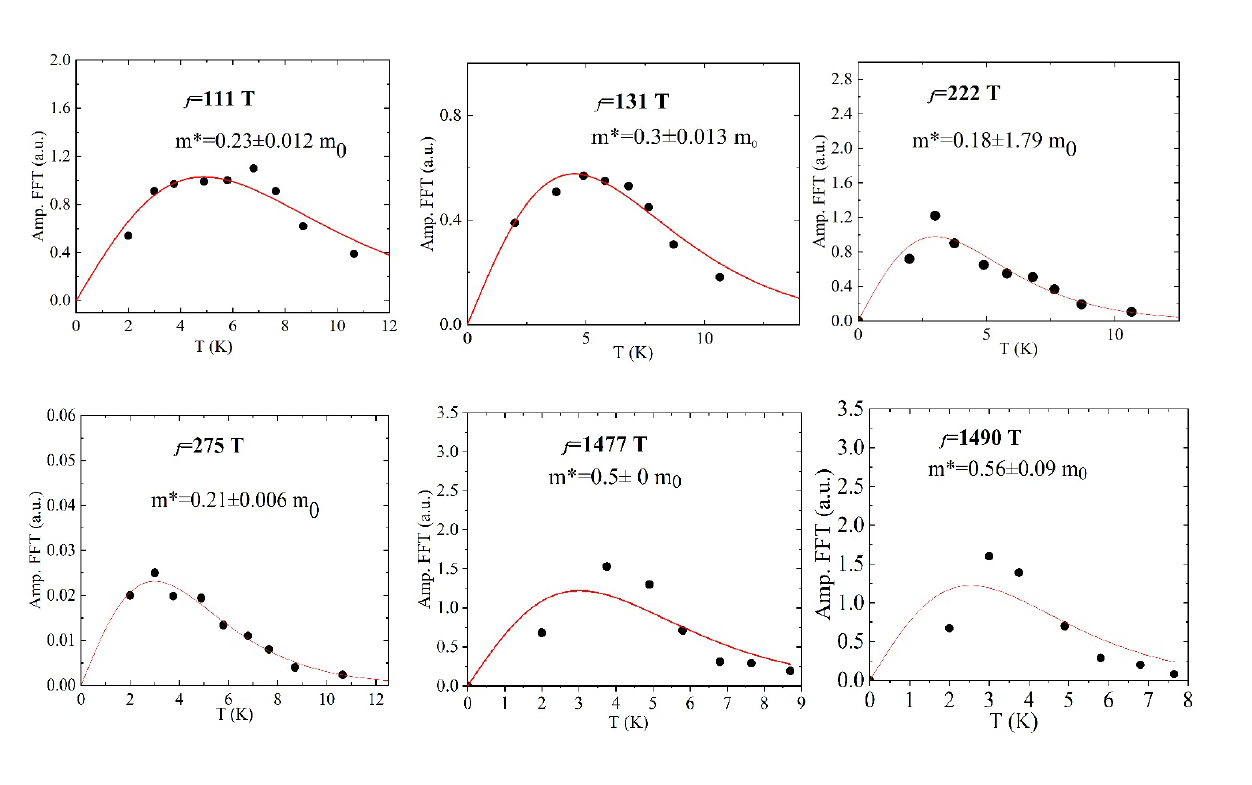}
  \caption {The peak heights determined from the corresponding FFT plots as a function of temperature for different frequencies. Solid red lines represent the effective mass fitting to equation (2).}
  \label{LKfits}
 \end{figure*}

With decreasing temperature, the high frequency oscillations become more pronounced. As shown in Fig.~\ref{FFTpeaks}, $F_{\alpha1}$ overlaps with $F_{\alpha2}$, which is almost a factor of three higher in amplitude. The different frequencies also differ in their temperature and field dependence. Fig.~5 presents the FFT peak amplitudes as a function of temperature for different orbit frequencies. The temperature dependences are notably different from the typical Lifshitz–Kosevich behavior observed in Shubnikov-de Haas (i.e. resistivity) and de Haas-van Alphen (i.e., magnetization) data, which exhibit a saturation at a maximum amplitude at the lowest temperatures \cite{Schoenberg}. 
According to the Mott formula \cite{Mott1958}, $S$ is related to the logarithmic derivative of the electrical conductivity $\sigma$ with respect to energy $\epsilon$ at the chemical potential $\mu$ by
\begin{equation}
\label{TEP}
    S=-\frac{\pi^2}{3}\frac{k_B^2T}{e} \left (\frac{\partial \rm{ln}(\epsilon)}{\partial \epsilon} \right)_\mu,
\end{equation}

Because TEP depends on the derivative of the density of states at the Fermi level and is related to the entropy carried by quasiparticles and their heat capacity, it is expected to vanish in the zero temperature limit.
Therefore the Lifshitz–Kosevich form applicable to quantum oscillations of TEP takes on a different form than the usual saturating behaviour found in other transport and thermodynamic quantitites.
The TEP amplitude is therefore given by \cite{morales2016thermoelectric},
\begin{equation}
\label{LK}
  A(T)\propto{{(\alpha pX)\coth (\alpha pX)-1}\over \sinh (\alpha pX)},
\end{equation}
which exhibits a maximum at $T\approx 0.11B / pm^*$ and approaches zero at $T\rightarrow 0$~K.\hspace {0.1 cm} Here, \hspace {0.1 cm} $\alpha=(2\pi^2k_B)/{e\hbar}$, where $k_B$ is the Boltzmann constant, $e$ is the elementary charge, $\hbar$ is the reduced Planck constant. $X=m^* T/B$ and $p$ is the harmonic number. To determine the cyclotron masses $m^*$ corresponding to each band, we have analyzed FFT amplitude as a function of temperature and fitted the data to Eq.~\ref{LK} as shown in Fig.~\ref{LKfits}. The effective mass of higher frequency bands (F$_{\alpha5}$, F$_{\alpha6}$) is half of $m_0$, whereas for the other frequencies the effective masses are similar of $0.2m_0$. The values obtained here for different frequencies are in good agreement with previously reported results from Shubnikov-de Haas measurements \cite{wang2020crystalline}. The lightest charge carriers originate from the $F_{\alpha1}$, $F_{\alpha2}$, and $F_{\alpha3}$ and $F_{\alpha4}$ bands. Below we investigate whether the phases of these oscillations exhibit non-trivial properties corresponding to Dirac fermions.


Zeeman splitting becomes observable in quantum oscillations at high magnetic fields, where the zeeman energy $E_z = g \mu_BB$ increases linearly with field and eventually exceeds the thermal and Dingle broadening of Landau levels. Although the ratio $\frac{E_z}{\hslash \omega_c}$ remains constant with field, the absolute energy separation between spin-split Landau levels grows, enabling resolution of spin-split features. This manifests as beating patterns, peak splitting, or the appearance of two closely spaced frequencies in Shubnikov-de Haas (SdH) or de Haas-van Alphen (dHvA) oscillations as observed in systems such as elemental Bismuth \cite{Zeeman-Bismuth}. 

\begin{figure}[h!]
  \centering
  \includegraphics[width=90mm]{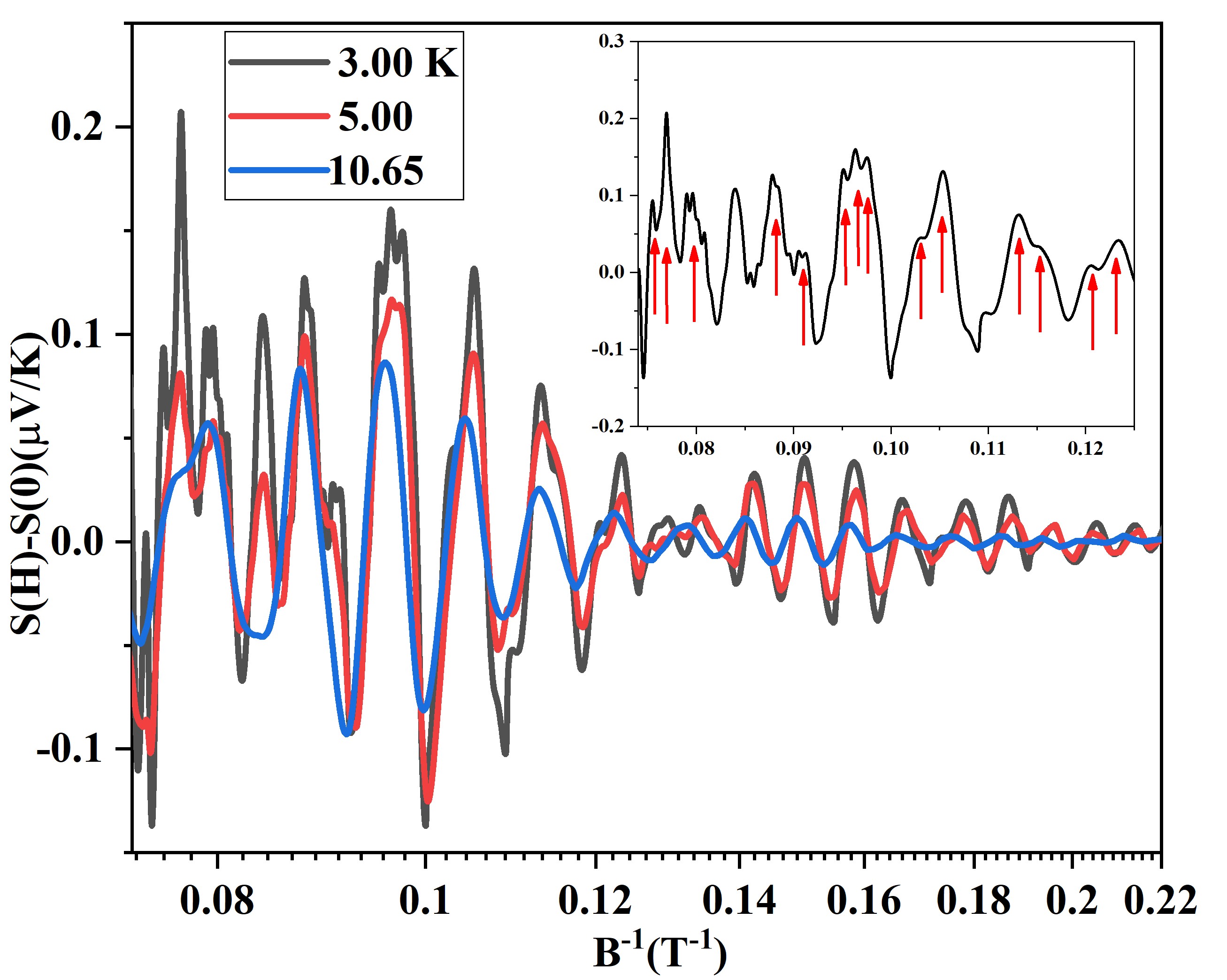}
   \caption{The oscillatory component of the magnetic field dependences of the thermoelectric power in BaAl$_4$ at selected temperatures. The split peaks are observed to gradually merge into single peak with raising temperature. The inset shows the peak splitting of Zeeman effect at T = 3 K as indicated by arrows.} 
   \label{split}
\end{figure}

For the frequency F$_{\alpha 4}$, the thermoelectric oscillations exhibit a strong Zeeman splitting starting at about 8 T. The threshold field for the peak splitting is 8 T which is low in comparison to the other well-known topological semimetals like Cd$_3$As$_2$ and TaP \cite{narayanan2015linear, hu2016pi}. As shown in Fig.~\ref{split}, the splitting of the oscillation peaks appears near 3 K (shown in inset) and the splitting of the LLs becomes less resolved with increasing temperatures, where the split peaks gradually merge into single peaks (Fig.~\ref{split}). This is consistent with the general expectation for the Zeeman effect. The Zeeman splitting also is observed in the quantum oscillation of magnetoresistance of BaAl$_4$ at 2 K 

At lower temperatures, the LLs begin to exhibit a doublet structure with a broad feature above 8 T, as shown in Fig.~\ref{split}. The separation of the doublet structure increases, when the magnetic field is increased, in particular the LL above 12 T completely evolves into two peaks. It indicates complete lifting of spin degeneracy due to the Zeeman effect. To analyze the Zeeman effect incurred at low temperatures, we calculate the energy difference between the spin split Fermi surfaces, $\Delta E=(gm^*)/(4m_0)=1/2(F/B_+ - F/B_- )$  (here $m^*$ is the cyclotron mass of charge carriers which depends on energy of the extremal Fermi surface areas which are proportional to the oscillation frequencies F and positive and negative sign represent the splitting of the peaks). The Landé $g$-factor is thus determined to be 10. 
  
\begin{figure}[h!]
  \centering
  \includegraphics[width=90mm]{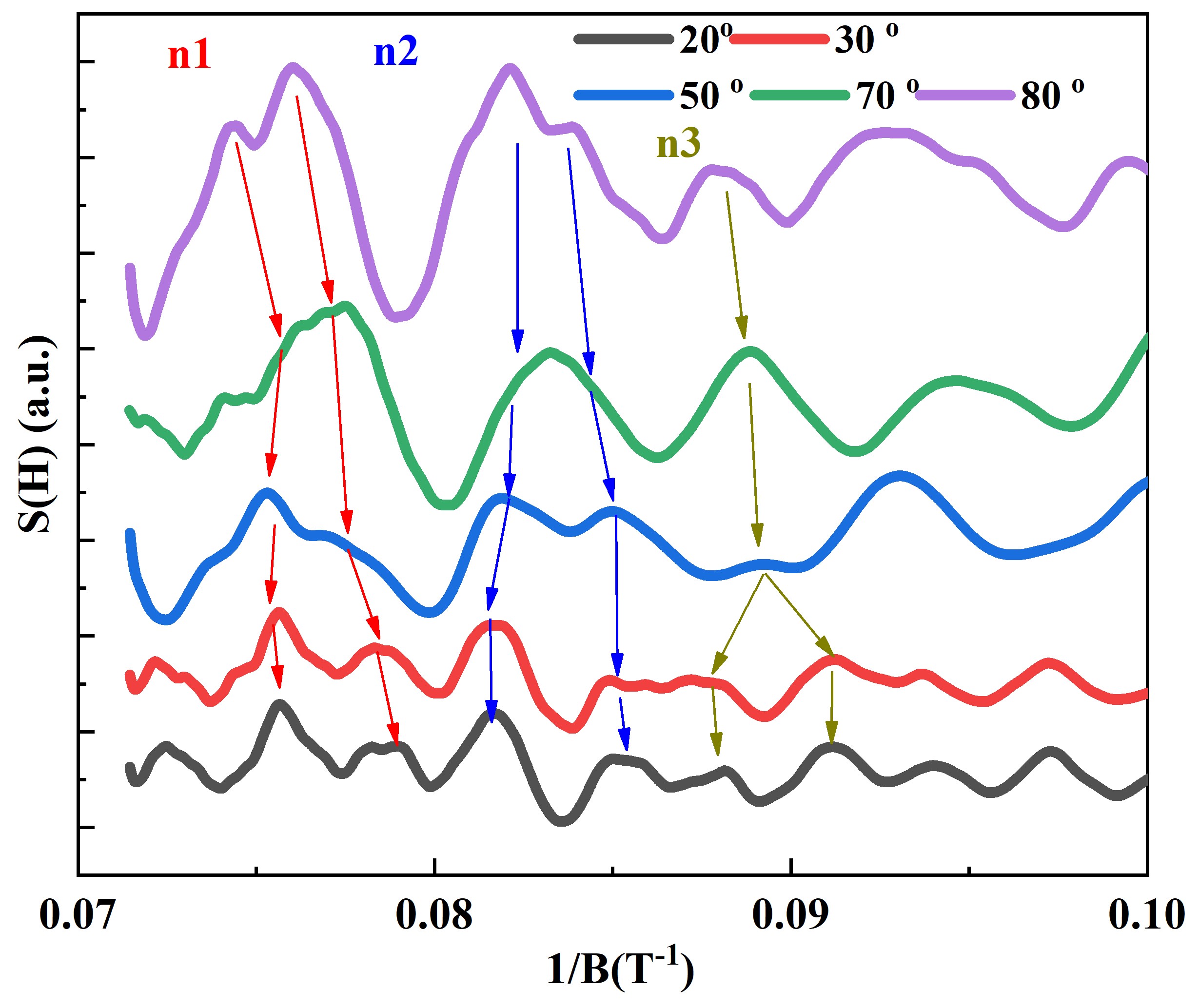}
  \caption{The angular dependence of thermopower oscillation at 2 K. The splitting of Landau levels in the high magnetic field are observed at different angles.} 
  \label{angle}
\end{figure}

The angular dependent Zeeman splitting is an important probe to understand the splitting mechanism i.e. to clarify the respective contribution from orbital and the Zeeman term in LL splitting. So we have performed the angle dependent thermopower measurements at higher magnetic field shown in Fig.~\ref{angle} to understand the origin of the Zeeman splitting. As shown in Fig.~\ref{angle} when the tilting angle $\theta$ is changed from $20^o$ to $80^o$ the splitting feature remains well resolved at all angles, but the spacing of the Zeeman splitting changes with the change of applied magnetic field angle. For BaAl$_4$, the thermopower oscillations are observed for both B//c and B//ab, which indicates that its electronic structure is 3D-like. In general, Zeeman splitting is scaled with the external magnetic field to make the spacing of the splitting LL unchanged with changing angle. If we consider the exchange interaction induced by the external field, the splitting may be decomposed into two parts, one orbital-dependent and another  orbital-independent. The orbital-dependent part mainly depends on the shape of the band structure leading to the angle-dependent splitting. In case of the orbital-independent part, the splitting does not change with change of angle and mainly depends on Zeeman effect \cite{wang2012dirac}. As shown in Fig.~\ref{angle}, the split spacing for all the peaks n1, n2, and n3 shows strong angular dependence denoting the large contribution from the orbital part. The orbital dependent splitting is observed to reach a maximum when the field is perpendicular to the [001] plane and vanishes when the field is parallel to the [001] directions. The strong orbital contribution in the Zeeman splitting may be caused by the spin-orbit coupling effect in BaAl$_4$ \cite{wang2020crystalline}. The orbital effect is significant in the a-c plane giving rise a highly anisotropic $g$-factor.

The band structure of topological Dirac semimetals hosts a nontrivial $\pi$ Berry phase which induces a phase shift in the quantum oscillations, which can be directly determined from the energy (i.e., inverse magnetic field) dependence of the LL index and is an important signature of topology. To look for such a phase shift in the high-frequency band, we plot the LL fan diagram (i.e., the LL index $n$ as a function of the inverse of magnetic field $1/B$) as shown in the Fig. 8. The index $n$ has been determined by considering the frequency $F$ obtained from FFT analysis and the positions of the peaks from $S$ versus $B^{-1}$ plots. We have excluded the fan diagram for higher frequencies because of the small number of observed peaks available for indexing. Fitting the LL index points linearly with two free parameters, shown in Fig. 8, yields slopes in n vs $B^{-1}$ plots is similar to the quantum oscillation frequencies determined from the FFT analysis. The phase shift $\phi_ T$ is then obtained from the intercepts, which are $n(0) =0.38\pm 0.04~ (\sim 3/8)$ for $F_{\alpha 2}$ and $n(0) = 0.9 \pm 0.08 ~(\sim 7/8)$ for $F_{\alpha 4}$.

So according to Eq.~\ref{TEP}, a phase shift of $\pi/2$ i.e. $\delta n= 1/4 $, is predicted to occur for diffusion thermopower oscillations relative to field-dependent resistivity oscillations.  The equivalent phase shift expected for SdH oscillation is $\phi=\phi_T\pm 1/4$, and the sign of the phase shift determines whether the orbit is electron-like or hole-like. The shift is $+\pi/2$ for electrons and $-\pi/2$ for holes \cite{matusiak2017thermoelectric}. The same sign change of the charge carriers is observed for graphene, for instance, where a change in phase is observed between thermoelectric power and resistivity \cite{zuev2009thermoelectric}. The phase shift $\phi$ in the case of SdH oscillations described by the Lifshitz-Kosevich formula is determined by the dimensionality of the Fermi surface, with $\pm 1/2$ and $\pm 5/8$ for 2D and 3D parabolic bands, respectively, and 0 and $\pm 1/8$ for 2D and 3D Dirac cones \cite{wang2016anomalous,mikitik1999manifestation, xiao2010berry}. So we can use the formula $\phi=\phi_T\pm 1/4$ (we are considering -ve sign, because from ARPES and quantum oscillation we have seen the hole pocket for these frequencies $F_{\alpha 2}$ and $F_{\alpha 4}$ \cite{wang2020crystalline}) for our thermoelectric data. For $F_{\alpha 2}$, $\phi=3/8-1/4=1/8$, which is expected to arise from 3D hole like Dirac cones. This phase shift of 1/8 also suggests a nontrivial Berry phase in BaAl$_4$ in the presence of a magnetic field which agrees with other studies \cite{liu2016zeeman, zheng2016transport}. This phase change is attributed to the change in Berry phase and hence non-trivial topology in the bands \cite{yuan2016observation}. Similarly for $F_{\alpha 4}$, the formula gives $\phi=7/8-1/4=5/8$. For a Dirac 3D topological semimetal, time-reversal symmetry leads to the discrete phase shift of $\phi=5/8$, as a function of the Fermi energy [29].
So we can summarize that we can ascribe the $F_{\alpha 2}$ band with a phase shift of 1/8, small effective mass, and a high mobility to the 3D Dirac band in BaAl$_4$, which is expected to display Dirac fermion properties confirming BaAl$_4$ as a topological semimetal with Dirac points. We also conclude that the $F_{\alpha 4}$ band shows a 3D Dirac cone. Both $F_{\alpha 2}$ and $F_{\alpha 4}$ bands indicate a 3D Dirac cone with non-trivial topology, which is consistent with the ARPES results and magnetoresistance quantum oscillation data \cite{wang2020crystalline}.

 \begin{figure} [h!]
 \centering
\includegraphics[width=90mm]{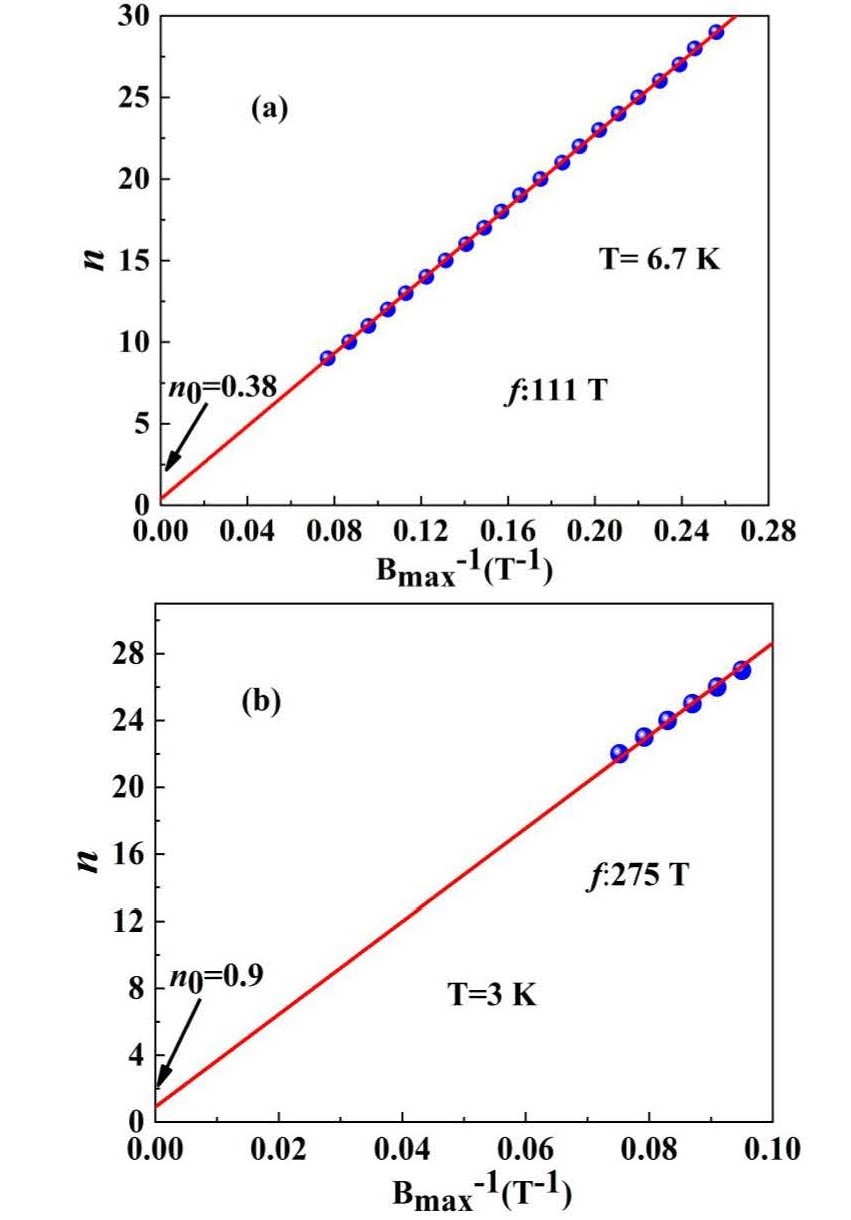}
    \caption{Landau-level (LL) index fan diagram for the frequencies 111 T ($F_{\alpha 2}$) and 275 T ($F_{\alpha 4}$). The red line represents the linear fit to all LL indices, which yields an intercept of $n_0$ = 0.38 for $F_{\alpha 2}$ and $n_0$ = 0.9 for $F_{\alpha 4}$.} 
    \label{}
    \end{figure}

\section{Conclusions}

In summary, we have observed strong thermoelectric quantum oscillations of the topological Dirac nodal-line fermions in BaAl$_4$. The analyses reveal six different oscillations with frequencies from $F_{\alpha 1}\approx 96$~T to $F_{\alpha 6}\approx 1490$~T. We also determine the corresponding effective masses by analyzing the Lifshitz–Kosevich amplitude dependence of thermoelectric power quantum oscillations. By determining the phase shifts we have observed that the $F_{\alpha 2}$ and $F_{\alpha 4}$ bands originate from 3D hole-like Dirac cones. Our study demonstrates that the orbital-dependent splitting is affected by the direction of the magnetic field and is significant in the a-c plane giving rise a highly anisotropic $g$-factor. 

 \section*{Acknowledgements}

Research at the University of Maryland was supported by the Gordon and Betty Moore Foundation’s EPiQS Initiative through Grant No. GBMF9071, the U.S. National Science Foundation Grant No. DMR2303090,
the Binational Science Foundation Grant No. 2022126, and the Maryland
Quantum Materials Center.

\newpage
\bibliographystyle{apsrev4-1} 
\bibliography{BaAL4}

\end{document}